\documentclass[journal]{IEEEtran}

\usepackage{graphicx}  

\makeatletter

\begin{document}

\title{Quantum Location Verification in Noisy Channels}

\author{Robert~A.~Malaney, School of Electrical Engineering and Telecommunications, University of New South Wales,  NSW 2052, Australia.
r.malaney@unsw.edu.au}
\maketitle

\begin{abstract}

 Recently it has been shown how the use of quantum entanglement can lead to the creation of real-time communication channels whose viability can be made location dependent. Such functionality leads to  new security paradigms that are not possible in classical communication networks. Key to these new security paradigms are quantum protocols that can unconditionally determine that a receiver is in fact at an \emph{a priori} assigned location.  A limiting factor of such quantum protocols  will be the decoherence of states held in quantum memory. Here we investigate  the performance of  quantum location verification protocols under decoherence effects. More specifically, we address the issue of how decoherence impacts the  verification  using
  $N=2$ qubits entangled as Bell states, as compared to  $N>2$ qubits entangled as GHZ states. We study the original quantum location
verification protocol, as well as a variant protocol, introduced here, which utilizes teleportation.
  We  find that  the performance of quantum location verification is in fact similar for Bell states and some $N>2$ GHZ states, even though
 quantum decoherence degrades  larger-qubit entanglements faster. Our results are important for the design and implementation of location-dependent communications in emerging quantum networks.

\end{abstract}

\section{Introduction}The ability to offer a real-time communication channel whose viability is unconditionally a function of the receiver location would offer a range of new information security paradigms and applications  (\emph{e.g.} see discussions in \cite{denning,classic1,malaney,classic}).  The ability to guarantee location-sensitive communications requires unconditional (independent of the physical resources held by an adversary) location verification. However, given that an adversary can possess unlimited receivers, each of which can be presumed to possess unlimited computational capacity, it is straightforward to see why classical-only unconditional location verification is impossible. Classical location verification use `challenges', and the finite speed of light, in order to bound ranges. As such, an adversary with multiple receivers can simply delay the response to  a challenge in order to circumvent verification (see discussion in  \cite{classic1}, \cite{classic} for more detail on verification limits of classical systems). However, recently it has been shown how the introduction of quantum entanglement into the communication channel leads to an unconditional \emph{quantum location verification} (QLV) protocol \cite{malaney2}. In \cite{malaney2} quantum entanglement is used to develop a `cloaked', and distributed, superdense coding system, in which the response times to (classical) challenges are measured in order to provide location verification.
 In QLV the answer to a classical challenge is encoded in entangled quantum states.
 Due to the fact that quantum information cannot be perfectly copied \cite{noclon}, only a device containing all qubits of the entangled state can decode successfully, and only a device at a specified location can answer within the required timescale. For a more  detailed description of QLV the reader is referred to \cite{malaney2}. For more details on superdense coding, including experimental status, the reader is referred to \cite{dense1}, \cite{dense6}. Details relating to some recent experiments involving entanglement, including its use in quantum teleportation \cite{tele1}, can be found in \cite{tenrife}, \cite{swap}. Discussion of the largest quantum communications network yet built can be found in  \cite{trusted}.

In \cite{malaney2} QLV was explored under the assumption of perfect quantum channels. More specifically, it was assumed that the principal resource of QLV, the entanglement between qubits, was preserved perfectly. In any practical system of course this will not be true. This is especially the case since  quantum memory is required for the most useful variants of the  QLV protocol. The current experimental status regarding quantum memory is reviewed in \cite{quantmemory} (see also \cite{quantmemory2}), where we learn that current timescales for successful storage of  multipartite entangled states is in the order of milliseconds. Heroic efforts to increase such storage times  to that required for workable large-scale quantum networks (\emph{i.e.} seconds) is currently underway in many laboratories \cite{quantmemory}. The limiting factor in quantum memory is decoherence effects, where qubit-environment interactions destroy the fragile entanglement of the quantum states. It is the purpose of this work to explore the effect quantum decoherence has on the performance of QLV. We will be specifically interested in the impact of decoherence on two-qubit maximally entangled states,  relative to three-qubit (and higher) maximally entangled states. As described more below,   QLV protocols use two-qubit and larger-qubit entangled states in a  different manner.

\section{Decoherence in QLV}
\subsection{Bell States and GHZ States in QLV}
Consider some system state $\left| {\Psi _s} \right\rangle$, which we can use as a means of encoding and decoding a secret sequence of bits. In the QLV protocol of \cite{malaney2} location verification of a device can be obtained by setting $\left| {\Psi _s} \right\rangle$ to the  Bell states.  If $N$ is the number of qubits entangled, the four  orthogonal basis states for a Bell state ($N=2$) can be written,
\[\left| {\Psi _s^{Bell}} \right\rangle  = \frac{1}{{\sqrt 2 }}\left( {\left| {00} \right\rangle  \pm \left| {11} \right\rangle } \right),\frac{1}{{\sqrt 2 }}\left( {\left| {10} \right\rangle  \pm \left| {01} \right\rangle } \right). \]
Alternatively,
Green-Horne-Zeilinger (GHZ)
\cite{ghz1} states in which $N=3$ qubits are maximally
entangled could be utilized. The eight  orthogonal basis states of a $N=3$ GHZ state can be written,
\[\left| {\Psi _s^{GHZ}} \right\rangle_{N=3}  = \left\{ \begin{array}{l}
 \frac{1}{{\sqrt 2 }}\left( {\left| {000} \right\rangle  \pm \left| {111} \right\rangle } \right),\frac{1}{{\sqrt 2 }}\left( {\left| {001} \right\rangle  \pm \left| {110} \right\rangle } \right) \\
 \frac{1}{{\sqrt 2 }}\left( {\left| {010} \right\rangle  \pm \left| {101} \right\rangle } \right),\frac{1}{{\sqrt 2 }}\left( {\left| {011} \right\rangle  \pm \left| {100} \right\rangle } \right) \\
 \end{array} \right\} .\]
  Larger qubit  GHZ states are also available. For example, a basis state of $N$ entangled qubits would be
 ${\left| {\Psi _s^{GHZ}} \right\rangle _N} = \frac{1}{{\sqrt 2 }}\left( {\left| {{{000...}_N}} \right\rangle  + \left| {{{111....}_N}} \right\rangle } \right)$
where the notation $\left| {{{000...}_N}} \right\rangle  = {\left| 0 \right\rangle ^{ \otimes N}}$
 is the tensor product of the state $\left| 0 \right\rangle $,
 $N$ times.\footnote{Creation of GHZ states is reviewed in \cite{wei}, where a ten-qubit \emph{hyper-entangled} state is also demonstrated. A QLV protocol based on hyper-entanglement would be an extension of the protocol presented in \cite{malaney2}.} Henceforth, a  state of the form  $\frac{1}{{\sqrt 2 }}\left( {\left| {{{000...}_N}} \right\rangle  + \left| {{{111....}_N}} \right\rangle } \right)$ is referred to as a `cat' state.

  In two dimensional location verification, at least three reference stations are required. With three reference stations, two Bell states would be required for each  \emph{instance of location verification}. Such an instance occurs every time the minimal amount of decoded information required for an independent location verification becomes available (\emph{e.g.} steps 4-6 of the protocol detailed later represent one such instance). The qubits of the two Bell states required for each instance of location verification would be distributed between the three reference stations (a single qubit at two of the stations, and two qubits at a third station). However, using GHZ states, only one  $\left| {\Psi _s^{GHZ}} \right\rangle_{N=3}$ state is required for the same instance of location verification. The three qubits of this state would also be distributed between the three reference stations (one qubit at each station).

For the protocol of \cite{malaney2}, if the number of  reference stations was also equal to $N$, then $\lceil N/2\rceil$ Bell states\footnote{$\lceil x \rceil$ is the ceiling of $x$.} will be required for each instance of location verification, as opposed to one $N$-qubit GHZ state.
 The question we now address is the following.
 What is the performance of a QLV protocol with $N$ reference stations, determined using $\lceil N/2\rceil$ Bell states, relative to the performance based on a single maximally entangled GHZ state comprising $N$ qubits? This question has important ramifications for the design of QLV protocols in emerging large-scale quantum networks.

\subsection{Decoherence Models}
 A decoherence model is built by studying the time evolution of some initial system state $\left| {{\Psi _s}} \right\rangle $, as its interacts with some external environment whose initial state we write as $\left| {{\Psi _e}} \right\rangle $. Without loss of generality we will assume $\left| {{\Psi _s}} \right\rangle $ and $\left| {{\Psi _e}} \right\rangle $ are initially not entangled with each other.

 In terms of the  density operators ${\rho _s} = \left| {{\Psi _s}} \right\rangle \left\langle {{\Psi _s}} \right|$ and ${\rho _e} = \left| {{\Psi _e}} \right\rangle \left\langle {{\Psi _e}} \right|$, the initial state of the combined total system can be written
 as ${\rho _s} \otimes {\rho _e}$. Although the open evolution of the system  ${\rho _s}$  is described by a non-unitary evolution, the closed evolution of ${\rho _s} \otimes {\rho _e}$ can be described by a unitary $U$ via ${U}({\rho _s} \otimes {\rho _e})U^\dag $. To obtain the output system state, $\rho _s^{out}$, after some closed evolution $U$, we use $\rho _s^{out} \equiv \varepsilon \left( {{\rho _s}} \right) = {\rm{T}}{{\rm{r}}_e}\left[ {{U}({\rho _s} \otimes {\rho _e})U^\dag } \right]$
where Tr$_e$ is the partial trace over the environment's qubits. The channel $\rho _s^{out} \equiv \varepsilon \left( {{\rho _s}} \right)$ is a completely positive, trace preserving, map which provides the required evolution of ${\rho _s}$. It is possible to describe such maps directly using  an operator-sum representation,
\begin{equation}
 \varepsilon \left( {{\rho _s}} \right) = \sum\limits_{a = 1}^M {{K_a}} {\rho _s}K_a^\dag ,{\rm{ \ where \ }}\sum\limits_{a = 1}^M {K_a^\dag {K_a} = I}  , \label{eqKraus}
\end{equation}
and where ${K_{a = 1...M}}$ represent the so-called Kraus operators, with $M$ being the number of operators \cite{Kraus}. One can show that a map given by Eq.~(\ref{eqKraus}) leads
 to affine transformations in the Bloch sphere coordinates of the state $\left| {{\Psi _s}} \right\rangle $ whose most general description requires 12  parameters. However, although  a general model of decoherence for a single qubit requires these 12 parameters, it is important to note that these parameters cannot be arbitrarily chosen due to the constraint of complete positivity (any map given by Eq.~(\ref{eqKraus}) is automatically completely positive). As we discuss more below, there are special-case quantum channels where only a few parameters are needed in order to map the decoherence of a qubit.

 Given a set of Kraus operators for a quantum channel, it is  straightforward to calculate the probability that, in some post-decoherence measurement, the state $\rho_s$ is  recovered. This probability, which we refer to as the fidelity $F$, is given by
 \begin{equation}
  F= {\rm{Tr}}(\rho_s \rho _s^{out}) = {\rm{Tr}}\left( {\rho_s \sum\limits_{a = 1}^M {{K_a}} {\rho _s}K_a^\dag } \right) .
 \label{pin}
\end{equation}
Note that $F$ as defined here is the square of the normal fidelity described by  ${\rm{Tr}}\sqrt {({\rho_s^{1/2}}\rho _s^{out}{\rho_s^{1/2}})} $. Clearly, a critical step is the determination  of the appropriate Kraus operators for a given channel.   As we are interested in probing QLV performance in the general case we will, in the first instance, construct the Kraus operators for the single qubits using random unitary matrices $U_a$. It is straightforward to show that given a set $\left\{ {{U_a}} \right\}$, and a set of real non-negative numbers $\left\{ {{p_a}} \right\}$
such that $\sum\nolimits_a {{p_a}}  = 1$, one can construct a set of Kraus operators $\left\{ {{K_a}} \right\}$ where ${K_a} = \sqrt {{p_a}} {U_a}$ (e.g. \cite{Nakahara}). For a single qubit there can be at most four ($a=1...4$) independent Kraus operators. Here we will construct four Kraus operators for a single qubit, by taking $U_2$, $U_3$ and $U_4$ to be random 2 dimensional unitaries. The first Kraus operator is set to the identity matrix ($U_1=I_2$). In addition, the probability $p_1$  associated with $U_1$ will be set equal to the decoherence parameter, $p$, where ${p} = 1 - {e^{ - \gamma t}}$, and where $t$ is the time spent in the channel and $\gamma$ is a rate associated with decoherence.
 By constraining the sets $\{p_a\}$ and $\{U_a\}$, it is possible to construct many different noisy channels.
 Although, in general, not mapping to any specific model of the qubit-environment interaction, the quantum channels just described, which we henceforth refer to as a \emph{random noise channels}, will allow us to investigate in a generic manner the \emph{relative} decoherence between the  $N=2$ (bipartite) Bell states and the $N>2$ (multipartite) GHZ entangled states. We ignore in this work, the (in-principle) possibility of  reversing decoherence in  random noise channels using classical information extracted from the environment \cite{GW3}.

 There are also decoherence channels modeled on specific qubit-environment interactions  (e.g. see \cite{chang}). For example, consider the depolarization channel.
 Using the following relations  for the Pauli matrices in the single-qubit basis; ${\sigma _o} = \left| 0 \right\rangle \left\langle 0 \right| + \left| 1 \right\rangle \left\langle 1 \right|$, ${\sigma _x} = \left| 0 \right\rangle \left\langle 1 \right| + \left| 1 \right\rangle \left\langle 0 \right|$, ${\sigma _y} = i(\left| 1 \right\rangle \left\langle 0 \right| - \left| 0 \right\rangle \left\langle 1 \right|$), and ${\sigma _z} = \left| 0 \right\rangle \left\langle 0 \right| - \left| 1 \right\rangle \left\langle 1 \right|$; we can write the density operator for the ${\left| {\Psi _s^{GHZ}} \right\rangle _N}$
 cat state as,
\begin{equation}
 \rho _s^{GHZ}(N) = \frac{1}{{{2^{N + 1}}}}\left( \begin{array}{l}
 {\left( {{\sigma _o} + {\sigma _z}} \right)^{ \otimes N}} + {\left( {{\sigma _o} - {\sigma _z}} \right)^{ \otimes N}} \\
  + {\left( {{\sigma _x} + i{\sigma _y}} \right)^{ \otimes N}} + {\left( {{\sigma _x} - i{\sigma _y}} \right)^{ \otimes N}} \\
 \end{array} \right). \label{eq:Kraus}
\end{equation}
We again introduce the  decoherence parameter, $p$,  of a qubit where $0 \le p \le 1$, with $p=1$ meaning complete decoherence and  $p=0$ meaning no decoherence.
 The depolarization channel for a single qubit can defined as $\varepsilon \left( {{\rho _s}} \right) = (1 - p){\rho _s} + p\frac{{{\sigma _o}}}{2}$. Using the relation \[{\sigma _o} = \frac{1}{2}\left( {{\rho _s} + \sum\limits_{j = 1}^3 {{\sigma _j}{\rho _s}{\sigma _j}} } \right),\] we see that the Kraus operators for the depolarization channel can be written
${K_1} = \sqrt {1 - \frac{{3p}}{4}} {\sigma _o}$, ${K_2} = \sqrt {\frac{p}{4}} {\sigma _x}$, ${K_3} = \sqrt {\frac{p}{4}} {\sigma _y}$, and ${K_4} = \sqrt {\frac{p}{4}} {\sigma _z}$.  Writing explicitly in terms of the Pauli matrices, and adopting time dependence $p = 1 - {e^{ - {\gamma _d}t}}$ we have the following model for a $N$-qubit GHZ state undergoing decoherence in the depolarization channel, ${\varepsilon _D}\left( {{\rho _s^{GHZ}}} \right)$
\begin{equation}
= \frac{1}{{{2^{N + 1}}}}\left( \begin{array}{l}
 {\left( {{\sigma _o} + {e^{ - {\gamma _d}t}}{\sigma _z}} \right)^{ \otimes N}} + {\left( {{\sigma _o} - {e^{ - {\gamma _d}t}}{\sigma _z}} \right)^{ \otimes N}} \\
  + {e^{ - N{\gamma _d}t}}\left[ {{{\left( {{\sigma _x} + i{\sigma _y}} \right)}^{ \otimes N}} + {{\left( {{\sigma _x} - i{\sigma _y}} \right)}^{ \otimes N}}} \right] \\
 \end{array} \right) .
\label{eq:dep}
\end{equation}
  Here (and in all our models) we have assumed  that   in  multi-qubit systems each qubit evolves in an equal and  independent manner.

Similarly, we can describe the amplitude damping channel as
\[{\varepsilon _{AD}}\left( {{\rho _s}} \right) = \left( {\begin{array}{*{20}{c}}
   {a + pc} & {\left( {\sqrt {1 - p} } \right)b}  \\
   {\left( {\sqrt {1 - p} } \right){b^*}} & {\left( {1 - p} \right)c}  \\
\end{array}} \right),\]
where
we have used  ${\rho _s} = \left[ {\begin{array}{*{20}{c}}
   a & b  \\
   {{b^*}} & c  \\
\end{array}} \right]$. This leads to two  Kraus operators  of the form ${K_1} = \left[ {\begin{array}{*{20}{c}}
   1 & 0  \\
   0 & {\sqrt {1 - p} }  \\
\end{array}} \right]$ and ${K_2} = \left[ {\begin{array}{*{20}{c}}
   0 & {\sqrt p }  \\
   0 & 0  \\
\end{array}} \right]$. This, in turn, leads to an amplitude damping channel explicitly given as ${\varepsilon _{AD}}\left( {{\rho _s^{GHZ}}} \right)$
\begin{equation}
 = \frac{1}{{{2^{N + 1}}}}\left( \begin{array}{l}
 {\left( {{\sigma _o} + {\sigma _z}} \right)^{ \otimes N}} + {\left( {{\sigma _o} + \left( {1 - 2{e^{ - {\gamma _a}t}}} \right){\sigma _z}} \right)^{ \otimes N}} \\
  + {e^{ - \frac{{N{\gamma _a}t}}{2}}}\left[ {{{\left( {{\sigma _x} + i{\sigma _y}} \right)}^{ \otimes N}} + {{\left( {{\sigma _x} - i{\sigma _y}} \right)}^{ \otimes N}}} \right] \\
 \end{array} \right) .
\label{eq:ad}
\end{equation}

A third commonly used channel is the phase damping channel which can be described by \[{\varepsilon _{PD}}\left( {{\rho _s}} \right) = \left( {\begin{array}{*{20}{c}}
   a & {\left( {1 - p} \right)b}  \\
   {\left( {1 - p} \right){b^*}} & c  \\
\end{array}} \right),\]
 leading to Kraus operators ${K_1} = \left( {\begin{array}{*{20}{c}}
   {\sqrt {1 - p} } & 0  \\
   0 & {\sqrt {1 - p} }  \\
\end{array}} \right)$,
  ${K_2} = \left( {\begin{array}{*{20}{c}}
   {\sqrt p } & 0  \\
   0 & 0  \\
\end{array}} \right)$, and ${K_3} = \left( {\begin{array}{*{20}{c}}
   0 & 0  \\
   0 & {\sqrt p }  \\
\end{array}} \right)$.
 This leads to a phase damping channel explicitly given as ${\varepsilon _{PD}}\left( {{\rho _s^{GHZ}}} \right)$
\begin{equation}
=\frac{1}{{{2^{N + 1}}}}\left( \begin{array}{l}
 {\left( {{\sigma _o} + {\sigma _z}} \right)^{ \otimes N}} + {\left( {{\sigma _o} - {\sigma _z}} \right)^{ \otimes N}} \\
  + {e^{ - N{\gamma _p}t}}\left[ {{{\left( {{\sigma _x} + i{\sigma _y}} \right)}^{ \otimes N}} + {{\left( {{\sigma _x} - i{\sigma _y}} \right)}^{ \otimes N}}} \right] \\
 \end{array} \right).
 \label{eq:ap}
\end{equation}
Kraus operators of the form ${K_1} = \sqrt p {\sigma _o}$ and ${K_2} = \sqrt {1 - p} {\sigma _\alpha }$, with $\alpha=x,y,z$, lead to the bit flip, the bit-phase flip, and the phase flip channels, respectively \cite{chang}.

Of course with the introduction of additional parameters, more general damping models  are available,  For example, the most general qubit model, subject to constraints that the decoherence commutes with rotations around the  $\sigma_z$ axis, and is continually differentiable and time stationary, is the model of  \cite{caves} where
${\varepsilon _Z}\left( {{\rho _s^{GHZ}}} \right)$
\begin{equation}
=\frac{1}{{{2^{N + 1}}}}\left( \begin{array}{l}
 {\left( {{\sigma _o} + \left[ {{e^{ - {\gamma _1}t}} + \mu \left( {1 - {e^{ - {\gamma _1}t}}} \right)} \right]{\sigma _z}} \right)^{ \otimes N}} \\
  + {\left( {{\sigma _o} - \left[ {{e^{ - {\gamma _1}t}} - \mu \left( {1 - {e^{ - {\gamma _1}t}}} \right)} \right]{\sigma _z}} \right)^{ \otimes N}} \\
  + {e^{ - N\left( {{\gamma _2} + i\omega } \right)t}}{\left( {{\sigma _x} + i{\sigma _y}} \right)^{ \otimes N}} \\
  + {e^{ - N\left( {{\gamma _2} - i\omega } \right)t}}{\left( {{\sigma _x} - i{\sigma _y}} \right)^{ \otimes N}} \\
 \end{array} \right),
 \label{eq:caves}
\end{equation}
where four real constants $\gamma_1$, $\gamma_2$, $\mu$, and $\omega$ are introduced.
Some of our previous decoherence models can be seen as special cases of this more general model. For example, setting $\mu  =  1$, ${\gamma _2} = {\gamma _1/2} = {\gamma _a}/2$, and $\omega  = 0$  in Eq.~(\ref{eq:caves}) leads to the amplitude damping channel of Eq.~(\ref{eq:ad}); and setting  ${\gamma _2}  = {\gamma _p}$, and ${\gamma _1}=  \omega =0$   in Eq.~(\ref{eq:caves}) leads to the phase damping channel of Eq.~(\ref{eq:ap}).

Combinations of random noise  channels and specific damping channels are also possible. For example, consider Krauss operators  ${K_1} = \sqrt{(1-\varepsilon_1)}\left( {\begin{array}{*{20}{c}}
   1 & 0  \\
   0 & {\sqrt {1 - p} }  \\
\end{array}} \right)$, ${K_2} = \sqrt{(1-\varepsilon_1)}\left( {\begin{array}{*{20}{c}}
   0 & {\sqrt p }  \\
   0 & 0  \\
\end{array}} \right)$
 $K_3=\sqrt{\varepsilon_3} U_3$, and $K_4=\sqrt{\varepsilon_4} U_4$, where  $\varepsilon_1$, $\varepsilon_3$, and $\varepsilon_4$ are in the range $0 - 1$, and where  $\varepsilon_1=\varepsilon_3+\varepsilon_4$. These operators lead to a quantum channel which approaches amplitude damping as $\varepsilon_1  \to 0$. In addition, if we let $U_3=I$ and set $\varepsilon_3=1-e^{-\gamma t}$, for some $\gamma$, this allows for a time dependence to be inserted into the additional random component of the channel.

\section{Results and Discussion}
\begin{figure}
 \includegraphics[width=3.5in,height=3.4in,clip,keepaspectratio]{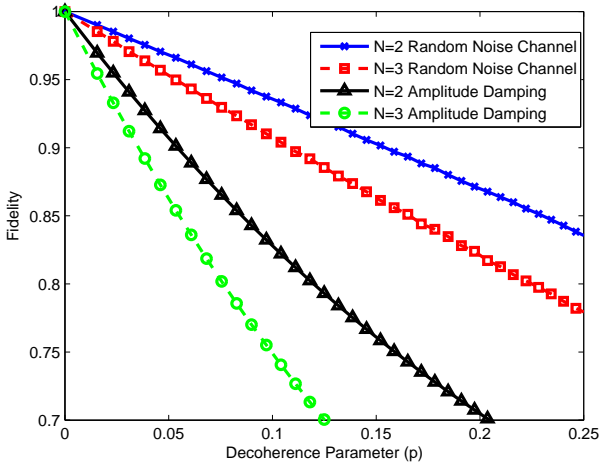}\\
  \caption{Fidelity vs. Decoherence Parameter ($p$) for random noise channels and amplitude damping, for Bell states and $N=3$ GHZ states.}
  \label{fig1}
\end{figure}

\begin{figure}
 \includegraphics[width=3.5in,height=3.4in,clip,keepaspectratio]{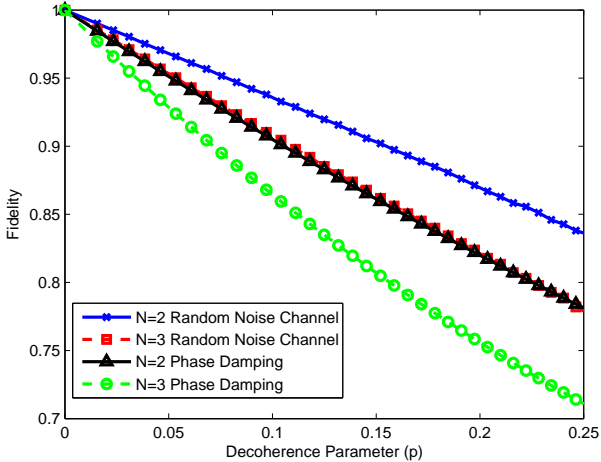}\\
  \caption{Fidelity vs. Decoherence Parameter ($p$), for random noise channels and phase damping, for Bell states and $N=3$ GHZ states.}
  \label{fig2}
\end{figure}

\begin{figure}
 \includegraphics[width=3.5in,height=3.4in,clip,keepaspectratio]{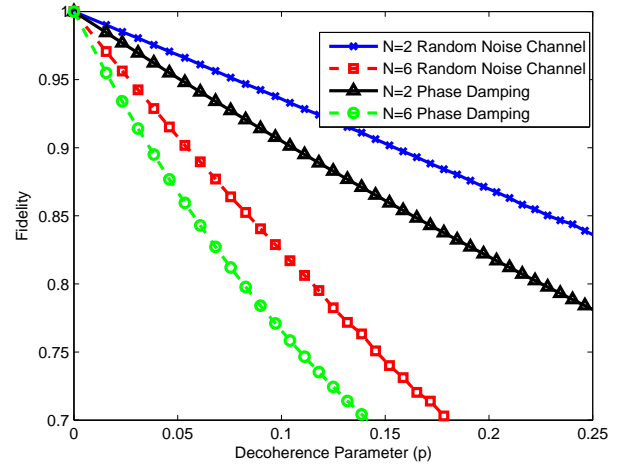}\\
  \caption{Fidelity vs. Decoherence Parameter ($p$), for random noise channels and phase damping, for Bell states and $N=6$ GHZ states.}
  \label{fig3}
\end{figure}

\begin{figure}
 \includegraphics[width=3.5in,height=3.4in,clip,keepaspectratio]{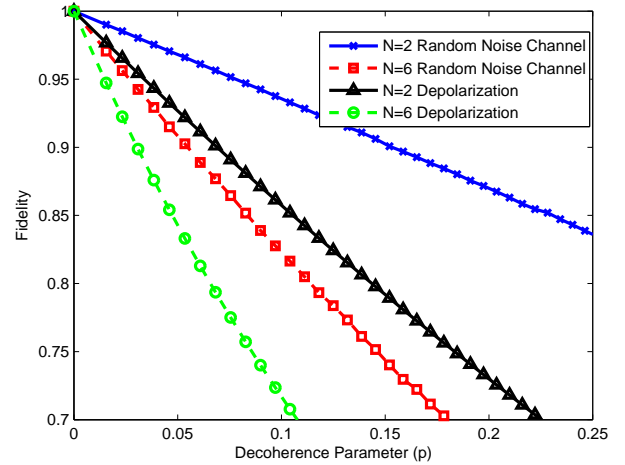}\\
  \caption{Fidelity vs. Decoherence Parameter ($p$), for random noise channels and depolarization, for Bell states and $N=6$ GHZ states.}
  \label{fig4}
\end{figure}

We have carried out many simulations over all of the decoherence channels described above for a wide range of parameter settings. Some of these results are shown graphically in Figs.~(\ref{fig1})-(\ref{fig4}). For the specific random noise channel shown the fidelity given is the average over 10,000 trials. In these plots we have focussed on the high fidelity region, and used cat states for the initial states. For QLV to be functional and unconditional it is important that only states whose fidelities remain high are utilized. A security threat to QLV is the potential
ability of an adversary, who is in possession of an optimal cloning
machine, copying  and redistributing the partial copies of entangled qubits
to other devices. If cloning were exact   QLV would fail (see \cite {malaney2} for details).
 However, for optimal cloning  $F$ is known to be upper
bounded by $F
 \approx 0.7$ for bipartite entanglement and $F
 \approx 0.6$ for tripartite entanglement  \cite{clone2,clone3}. For  a series of two-bit messages encoded in $L=100$   Bell  states, an adversary with access to an optimal cloning machine would have a probability of 1 in
$10^{16}$ of passing
 the verification system even though not at the authorized location.
 Arbitrary smaller probabilities are achieved exponentially in $L$.  Within QLV, decoherence must be limited so as to provide for a decoded bit error rate significantly above that expected if an optimal cloning machine was present. Clearly, a value of $F=0.9$ would suffice in this regard, and we will focus on this value in the discussion of our results.

 Our results for the fidelity cannot be directly mapped to true time $t$, as this would require a detailed understanding of the decoherence rates of the channels. However, they can be used to determine the performance of Bell states relative to  $N$-qubit GHZ states within the QLV context. We can compare performance levels by determining the probability of an instance of location verification. As discussed earlier,
 an instance of location verification occurs when enough decoding has occurred for a single estimate of location to be obtained. In two dimensional space, timings from at least three reference stations are needed. This means if Bell states are used as the encoding states, then two Bell states must be used in the encoding. If a GHZ state is used, then only one
  ${\left| {\Psi _s^{GHZ}} \right\rangle _{N = 3}}$ state is required in the encoding. If, following decoherence, $F_B$ and  $F_{GHZ}$ are the fidelities of the Bell states and GHZ states, respectively, then the probability of a single instance of location verification for each case can be determined. This is achieved for the Bell state case with probability $F_B^2$, whereas for the GHZ case it would simply be $F_{GHZ}$.

  In Fig.~(\ref{fig1}) the fidelity as a function of $p$ is shown for the Bell state ($N=2$ curves) and the   ${\left| {\Psi _s^{GHZ}} \right\rangle _{N = 3}}$ state ($N=3$ curves), in both random noise channels and amplitude damping channels.  From this figure we see that a fidelity of $F=0.9$  is reached at $p\approx 0.07$ for Bell states in the amplitude damping channel.
We can also see for the $N=3$ GHZ state, the fidelity in the amplitude damping channel will be $\sim 0.85$ at the same $p$. Recall the probability of a successful instance of location verification for Bell states at $F=0.9$ will be 0.81.
  We can see, therefore, that the use of GHZ states will provide slightly better performance relative to Bell states  for the specific amplitude damping channel shown.  Similar arguments and conclusions also hold with regard to the random noise channels.
  A similar conclusion to above can be drawn from Fig.~(\ref{fig2}), where phase damping channels have been investigated.  Again, GHZ states perform slightly better than Bell states for a  range of  fidelities once the probability of a successful instance of location verification is calculated.

  In Fig.~(\ref{fig3}) the fidelity as a function of $p$ is shown for Bell state ($N=2 $ curves)  and the   ${\left| {\Psi _s^{GHZ}} \right\rangle _{N = 6}}$ state ($N=6$ curves) in both random noise channels and phase damping channels. Here we start to see the impact of the faster decoherence of higher dimensional entangled states. In this case, we are modeling the scenario where six reference stations are utilized in order to obtain an instance of location verification. Additional reference stations would increase the location accuracy of the verification. Note, we mean here by increased location accuracy, that achieved  by  increases in signal-to-noise, and  improved dilution of precision effects arising from extra reference stations.\footnote{Quantum meteorology effects (e.g. \cite{giov} \cite{caves}) can also provide increased location \emph{accuracy}. However, the additional accuracy obtained via quantum meteorology can also be obtained using additional classical resources (\emph{i.e.} without quantum entanglement). Unconditional location \emph{verification}  cannot be achieved using accuracy improvements offered by quantum meteorology, nor can it be achieved with the use of additional classical resources (\emph{i.e.} verification requires quantum entanglement).}

   Fig.~(\ref{fig3}) is useful for a discussion of the case when we have three Bell states deployed for each instance of location verification. This can be compared to a single decoding of the  ${\left| {\Psi _s^{GHZ}} \right\rangle _{N = 6}}$ state. For the Bell states, a target fidelity of 0.9 in the phase damping channel occurs at $p \sim 0.1$. The probability of decoding all three Bell states at this $p$ is then $\sim 0.73$. This is approximately the same probability of $0.75$ for decoding the ${\left| {\Psi _s^{GHZ}} \right\rangle _{N = 6}}$ state at the same  $p$. This  suggests $N=6$ is close to the limit where higher dimensional states  will retain a performance in QLV similar to that achieved with Bell states. Note, however, for Bell states we would need only 2 out of the 3 Bell states to  be successfully decoded in order to obtain an instance of location verification. At a target fidelity of 0.9 the probability of this outcome would be 0.97, so in this sense Bell states are operating more efficiently than  ${\left| {\Psi _s^{GHZ}} \right\rangle _{N = 6}}$ states. But the removal of one of the Bell states in the decoding would lead to a
   reduced location accuracy relative to 3 out of 3 Bell state decodings. As such, strictly speaking we should state that the performance (at high fidelity) of the Bell states and  ${\left| {\Psi _s^{GHZ}} \right\rangle _{N = 6}}$ states  is the same for an instance of location verification at \emph{equal location accuracies}.

    In Fig.~(\ref{fig4}) we  can again compare to single decoding of the  ${\left| {\Psi _s^{GHZ}} \right\rangle _{N = 6}}$ state to three decodings of Bell states, this time for
     the depolarization  channel of Eq.~(\ref{eq:dep}). The results shown in Fig.~(\ref{fig4}) lead us to a similar conclusion to that drawn previously. That  is, generally speaking, there is no significant performance degradation in QLV when a single GHZ state (with $N< 6$) is used  in place of multiple  Bell states. Although not shown, the constrained $\sigma_z$-rotation decoherence of Eq.~(\ref{eq:caves}), and damping channels combined with random noise contributions, all lead to a similar general conclusion.

\section{Entanglement-Swapping QLV}
  In   \cite{malaney2}, physical transfer of the qubits is undertaken. Here we introduce a variation to the protocol of \cite{malaney2} where physical transfer of qubits is replaced by teleportation of the qubits. The physical transfer is negated  by having the device (whose location is to be verified) possess stored qubits that are \emph{a priori} entangled with one reference station.\footnote{Such \emph{a priori} entanglement does not by itself produce location authentication since such entanglement can be re-distributed. Note also, the \emph{a priori} entanglement could be used for remote state preparation.} The `trade-off' is the requirement in the new protocol for quantum memory which can hold state information for much longer timescales relative to the protocol of \cite{malaney2}. This trade-off between physical transfer of qubits  and quantum memory requirements, will be of value as advances in quantum memory develop.
 For clarity we will present the new protocol for a one dimensional location verification using Bell states in which only two reference stations are used. Extensions to the two dimensional problem (three reference stations) are then discussed.

 Consider two  reference stations at publicly known locations, and
a device (Cliff) that is to be verified at a publicly known
location $(x_v,y_v)$.  We assume that the reference stations (Alice and Bob) are
authenticated
and share  secure communication channels between
each other via quantum key distribution
\cite{qcd1,qcd2}. We also assume that
 all  classical communications  between Cliff and the reference stations
   occur at the speed of light, $c$.
Processing time (\emph{e.g.} due to local quantum measurements) is
assumed negligible. Given these assumptions, we wish  to unconditionally verify Cliff is  at the location
$(x_v,y_v)$.

In QLV a geometrical constraint on the reference stations is always required. For
one-dimensional location verification the constraint is that $ \tau _{AC} +
\tau _{BC} = \tau _{AB} $, where $ \tau _{AC}$ ($ \tau _{BC}$) is
the light travel time between Alice (Bob) and Cliff, and where
$\tau_{AB}$ is the light travel time between Alice and Bob.
Let Alice share with
Cliff a set of \emph{L} entangled qubit pairs  (\emph{i.e.} a set of $L$ Bell states) $ \Omega _i \left[
{AC} \right]$, where the subscript $i = 1 \ldots L$ labels the
entangled pairs. We let the pairs be
labeled (\emph{e.g.} by memory address) in the order received from some source,
  with the
first  qubit of each pair being held by Alice  and the second by Cliff.
 We will
assume an encoding  ($ 00 \to
\frac{1}{{\sqrt 2 }}\left( {\left| {00} \right\rangle  + \left|
{11} \right\rangle } \right)$ \emph{etc.}) that is
public.
  Let Alice also share with Bob a different set of
\emph{L/2} entangled qubit pairs $\Lambda_j \left[ {AB} \right]$,
$j = 1 \ldots L/2$.  Without loss of generality we can
assume all entangled qubit pairs are initially in the state $\frac{1}{{\sqrt 2 }}\left( {\left| {00} \right\rangle  + \left|
{11} \right\rangle } \right)$.  An entanglement-swapping QLV protocol proceeds as follows.

Step 1: Alice initiates an  entanglement swapping procedure in
order to form  a new set $ \Gamma _j \left[ {{{BC}}} \right]$ of
$L/2$  entangled pairs between Bob and Cliff. She achieves this by
\emph{randomly selecting} one of her local qubits from the pairs
$\Omega _i \left[ {AC} \right]$, combining this with
 one of her local qubits sequentially chosen from the pairs $\Lambda_j \left[
{AB} \right]$, and conducting a Bell State Measurement (BSM) on
the two qubits. These qubits are not selected again for BSM. Alice
repeats this process until all of her local qubits from the pairs
$\Lambda_j \left[ {AB} \right]$ have undergone BSM. At this point
Bob shares a new set $\Gamma_j \left[ {BC} \right]$ of $L/2$
entangled pairs with Cliff, and Alice shares a reduced set $\Omega
_{j'} \left[ {AC} \right]$ of $L/2$ entangled pairs with Cliff
($j' = 1 \ldots L/2$). We label our sets with the different
subscripts $i,j,j'$ to illustrate the following points. Cliff is
in possession of $L$ qubits  which remain labeled with the index
$i$. He is unaware which reference station (Alice or Bob) each of
the qubits in his possession is entangled with.

Step 2: Alice communicates with Bob via their secured channel, and
informs him of two facts related to each of the local qubits he
possesses from the pairs $\Gamma_j \left[ {BC} \right]$. Bob is
informed of the BSM result relevant to each qubit, and  the $j \to
i$ mapping.

Step 3: Alice generates a random binary sequence $S_a$ of length
\emph{K} bits $(K<L)$, and shares this sequence with Bob. This sequence is encoded into a different set of Bell states $\Lambda'_j \left[ {AB} \right]$  \emph{a priori} shared by Alice and Bob. Who undertakes the  local unitary operation needed to encoded each two-bit segment of the sequence $S_a$ into $\Lambda'_j \left[ {AB} \right]$ is decided by Alice and Bob for each segment.

Step 4: Using a qubit, say qubit $j'$, from the set
$\Omega_{j'} \left[ {AC} \right]$, Alice undertakes a BSM with the first qubit she holds from the set   $\Lambda'_j \left[ {AB} \right]$. Using a classical channel she  informs Cliff  the outcome of the BSM and the label $i$ of the qubit held by Cliff
to which the BSM outcome relates to. This completes the teleportation of the qubit  $\Lambda'_1 \left[ {AB} \right]$ from Alice to Cliff. Likewise, Bob teleports the corresponding qubit of $\Lambda'_1 \left[ {AB} \right]$ held locally by him. The classical messages from Alice and Bob related to the teleportation  are sent so as to arrive at the location  $(x_v,y_v)$ simultaneously.

 Step 5: Using the classical information received from Alice and Bob, Cliff performs a BSM in order to decode
two bits of information. Cliff then \emph{immediately}
communicates classically to Alice and Bob informing them
of the two  bits  he decoded.

Step 6: Alice checks that the sequence returned to her by Cliff is
correctly decoded and notes the round-trip time for the process.
Likewise Bob. Alice and Bob can then compare their round-trip
times to Cliff (2$\tau_{AC}$ and 2$\tau_{BC}$) in order to verify
consistency with Cliff's publicly reported location $(x_v,y_v)$.

 Extension of the
above protocol  to two-dimensional
verification could be achieved by the introduction of additional Bell states and a third reference station, say Dan. Alice would repeat the procedures with Dan that she undertook with Bob. With such a set-up, teleportation of two Bell states would be required for an instance of location verification.
However, another
 solution  is the use of a $\left| {\Psi _s^{GHZ}} \right\rangle_{N=3}$ state.   We will not pursue here a detailed exposition of the protocol in this case, except to note the following. At some point in the protocol one qubit from the  $\left| {\Psi _s^{GHZ}} \right\rangle_{N=3}$ state would be held at each of three reference stations. The teleportation of the three qubits to Cliff would comprise an instance of location verification.

 The relative performance of two Bell states, as compared to a single $\left| {\Psi _s^{GHZ}} \right\rangle_{N=3}$ state, in the entanglement-swapping protocol just discussed,  would follow a similar discussion to that given in Section~III. However, we do note the additional  entanglement (needed for teleportation) required in the entanglement-swapping protocol  would also suffer decoherence. Although this extra decoherence will not directly impact the relative performance of Bell states as compared to a  $\left| {\Psi _s^{GHZ}} \right\rangle_{N=3}$ state, it will directly degrade (albeit slightly) the performance of the entanglement-swapping protocol relative to protocols that use direct transfer of qubits.

\section{Conclusions}

 We have investigated the relative impact of  decoherence on QLV protocols.
  We find that, in general, it is possible to verify locations  using multipartite GHZ states, at a comparable performance level  to that obtained  using multiple Bell states.
  Efforts at creating on-demand long-term quantum  memory are now bearing fruit, and quantum networks are currently being built and tested. Deployment of QLV  protocols in such networks will, for the first time, offer the opportunity to deliver real-time communications systems which can be made unconditionally dependent on the physical location of a receiver.


\end{document}